# INTRODUCING AIC MODEL AVERAGING IN ECOLOGICAL NICHE MODELING: A SINGLE-ALGORITHM MULTI-MODEL STRATEGY TO ACCOUNT FOR UNCERTAINTY IN SUITABILITY PREDICTIONS

ELIÉCER E. GUTIÉRREZ[1, 2] AND NEANDER M. HEMING[2]

[1] Programa de Pós-Graduação em Biodiversidade Animal, Centro de Ciências Naturais e Exatas, Av. Roraima n. 1000, Prédio 17, sala 1140- D, Universidade Federal de Santa Maria, Santa Maria, RS 97105-900, Brazil

[2] Programa de Pós-Graduação em Zoologia, Departamento de Zoologia, Universidade de Brasília, 70910-900 Brasília, DF, Brazil

## ABSTRACT

*Aim:* The Akaike information Criterion (AIC) is widely used science to make predictions about complex phenomena based on an entire set of models weighted by Akaike weights. This approach ("AIC model averaging"; hereafter AvgAICc) is often preferable than alternatives based on the selection of a single model. Surprisingly, AvgAICc has not yet been introduced in ecological niche modeling (ENM). We aimed to introduce AvgAICc in the context of ENM to serve both as an optimality criterion in analyses that tune-up model parameters and as a multi-model prediction strategy.

*Innovation:* Results from the AvgAICc approach differed from those of alternative approaches with respect to model complexity, contribution of environmental variables, and predicted amount and geographic location of suitable conditions for the focal species. Two theoretical properties of the AvgAICc approach might justify that future studies will prefer its use over alternative approaches: (1) it is not limited to make predictions based on a single model, but it also uses secondary models that might have important predictive power absent in a given single model favored by alternative optimality criteria; (2) it balances goodness of fit and model accuracy, this being of critical importance in applications of ENM that require model transference.

*Main conclusions:* Our introduction of the AvgAICc approach in ENM; its theoretical properties, which are expected to confer advantages over alternatives approaches; and the differences we found when comparing the AvgAICc approach with alternative ones;

should eventually lead to a wider use of the AvgAICc approach. Our work should also promote further methodological research comparing properties of the AvgAICc approach with respect to those of alternative procedures.

*Key words:* Akaike information criterion, ecological niche modeling, MaxEnt, model accuracy, model complexity, model transferability, model tuning, model uncertainty, species distribution modeling

## INTRODUCTION

Ecological niche modeling (ENM) embodies a set of tools used in areas of research ranging from ecology, evolution, and biogeography to conservation biology and public health. Largely responsible for this broad use is the possibility to transfer models onto environmental scenarios both cross-geography and cross-time. Paradoxically, model transferability is one of the aspects of ENM that requires most attention by researchers (Randin et al., 2006; Peterson et al., 2007; Anderson, 2012; Halvorsen et al., 2015).

It has been shown that model complexity plays a major role on model transferability (Warren & Seifert, 2011; Shcheglovitova & Anderson, 2013; Warren et al., 2014; Moreno-Amat et al., 2015). For instance, alternative, competing models might perform in a similar fashion when projected onto the environmental scenario in which they were calibrated; however, their behavior can substantially differ from each other once transferred to other scenarios. Once transferred, over-parameterized models tend to underestimate the availability of environmentally suitable conditions for the focal species, whereas under-parameterized models tend to overestimate it (Warren & Seifert, 2011). A recommendation emerging from studies dealing with the effect of model complexity on model transferability is to optimally balance complexity and accuracy of models through tuning of model parameters (Warren & Seifert, 2011; Shcheglovitova & Anderson, 2013; Radosavljevic & Anderson, 2014; Warren et al., 2014; Moreno-Amat et al., 2015; e.g., Boria et al., 2017; Galante et al., 2017). This is achieved through calibration of preliminary models built with varying values of model parameters. The complexity and accuracy of these preliminary models are then compared in order to select the set of parameters that yields the "best model". A final model is then calibrated with these chosen parameters.

What optimality criteria should be used to identify the "best model" among preliminary models? This is an open question. Regarding to the type of information that optimality criteria for model selection convey, they can be classified in two groups. The first group is based on metrics of model accuracy, i.e., metrics that assess the degree in which a model incurs in omission, commission, or both kinds of errors. Examples of these metrics are the Area Under the Curve of the Receiver Operating Characteristic plot (AUC) and omission rates (OR; see Peterson et al., 2011 and references therein). The second group measures goodness of fit, i.e., how well a model output matches the likelihood of the species' occurrence data used to estimate the parameters of the model. For tuning parameters of ecological niche models based on presence-only data, the Akaike Information Criterion (AIC) is the prevailing metric of goodness of fit.

Relative to other metrics, AIC has the advantage that it balances goodness of fit and model complexity, which, as aforementioned, is pivotal for model transferability. To date, AIC is used to select a single set of model parameters (i.e., those of the preliminary model with the lowest AIC score) for the calibration of a final model. However, because preliminary models occupying lower ranks in the hierarchy of AIC weights can provide information absent in the single model deemed to be the best one, this procedure might miss the opportunity to capitalize on useful information. An alternative approach is to use a procedure termed "model averaging", which employs all informative models to generate a weighted average prediction, with models contributing to the average prediction proportionally to their respective AIC weights (Burnham & Anderson, 2002). Model averaging is theoretically preferable to the use of a single model (deemed to be the "best model") because it employs a broader spectrum of predictive information, making it a promising strategy for ENM studies dealing with uncertainties associated to model transferability. Although consensus predictions have been implemented in ENM (Araújo & New, 2007; Buisson et al., 2010; Zhang et al., 2015), the use of AIC to select and objectively weight the contribution of individual models on the consensus prediction has not been employed. Hence, we herein introduce the AIC model averaging consensus approach (AvgAICc) in ENM both as an optimality criterion for tuning model parameters and as a multi-model predictive strategy. We regard AvgAICc as an optimality criterion because it does control how much each of the models considered will contribute to a final inference, with some of the models influencing such inference negligibly. Note that AIC averaging has been used in occupancy modeling for a completely different purpose, which is to average coefficients of predictors. Besides, occupancy modeling generates predictions of presence in geographic regions that are relatively small. By contrasts, ecological niche modeling generates predictions of suitability—often at a large geographic scale—that can (and often are) transferable geographically and temporarily. To the best of our knowledge, AIC model averaging to generate an assembled prediction has only been used in a single ecological niche modeling study (i.e., Moussalli et al., 2009), although its potential utility has been mentioned in the literature (e.g., Guisan and Thuiller 2005; Elith and Leathwick 2009). Surprisingly, the approach has not been formally introduced and comparisons with alternative methods for conducting both parameter tuning and assembling suitability models have not been carried out. To remedy the situation, we introduce the AvgAICc approach using the most popular algorithm (MaxEnt) and type of data (presence-only data) hoping that our work will serve as guidance for and encourage interest in further research that, based on a variety of (real and virtual) species and geographic and environmental scenarios, should test the potential of this approach. Meanwhile, we focus our efforts in describing the AvgAICc approach and addressing a simple question: Does the AvgAICc strategy favor models that differ from those favored by other commonly used optimality criteria (i.e., Lowest AICc score, omission rate, AUC) with regard to model complexity, importance of environmental variable, and amount and geographic locations of areas harboring suitable conditions?

## METHODS

The following paragraphs provide information regarding the methods herein employed; a much more detailed description of the methods used was placed in Appendix S1.

***Experimental design.***—We modeled an approximation to the fundamental niche (Soberón, 2007) of *Bradypus variegatus* based on bioclimatic variables and the software MaxEnt, and compared different model-tuning optimality criteria and prediction strategies. Through preliminary models, we tuned up MaxEnt's feature classes and the regularization multiplier, both of which heavily impact model complexity and model accuracy (Shcheglovitova & Anderson, 2013; Radosavljevic & Anderson, 2014). We employed four optimality criteria to select the (estimated) best sets of model parameters for the calibration of a final model (one per optimality criterion) and compared the behavior of the resulting final models. We emphasized comparisons of the AIC model averaging approach (AvgAICc, introduced herein) with alternative strategies regarding model complexity, importance of environmental variables, and amount and geographic location of areas predicted as suitable for the focal species.

*Species occurrences, environmental data, and study regions.*—The focal species, *Bradypus variegatus,* ranges from Honduras to northern Argentina, at elevations from sea level to above 2,400 meters (Fig. 1; Gardner, 2008; Hayssen, 2010). The species has not been subjected to modern taxonomic revision; hence, the results of the analyses presented here should serve as a basis for comparisons of model techniques. We employed 108 occurrences obtained from the R package "dismo" (Hijmans et al., 2017; Fig. 1; See additional files in Gutiérrez & Heming, 2018) that were separated by a distance of at least 10 km from each other (Appendix S1).

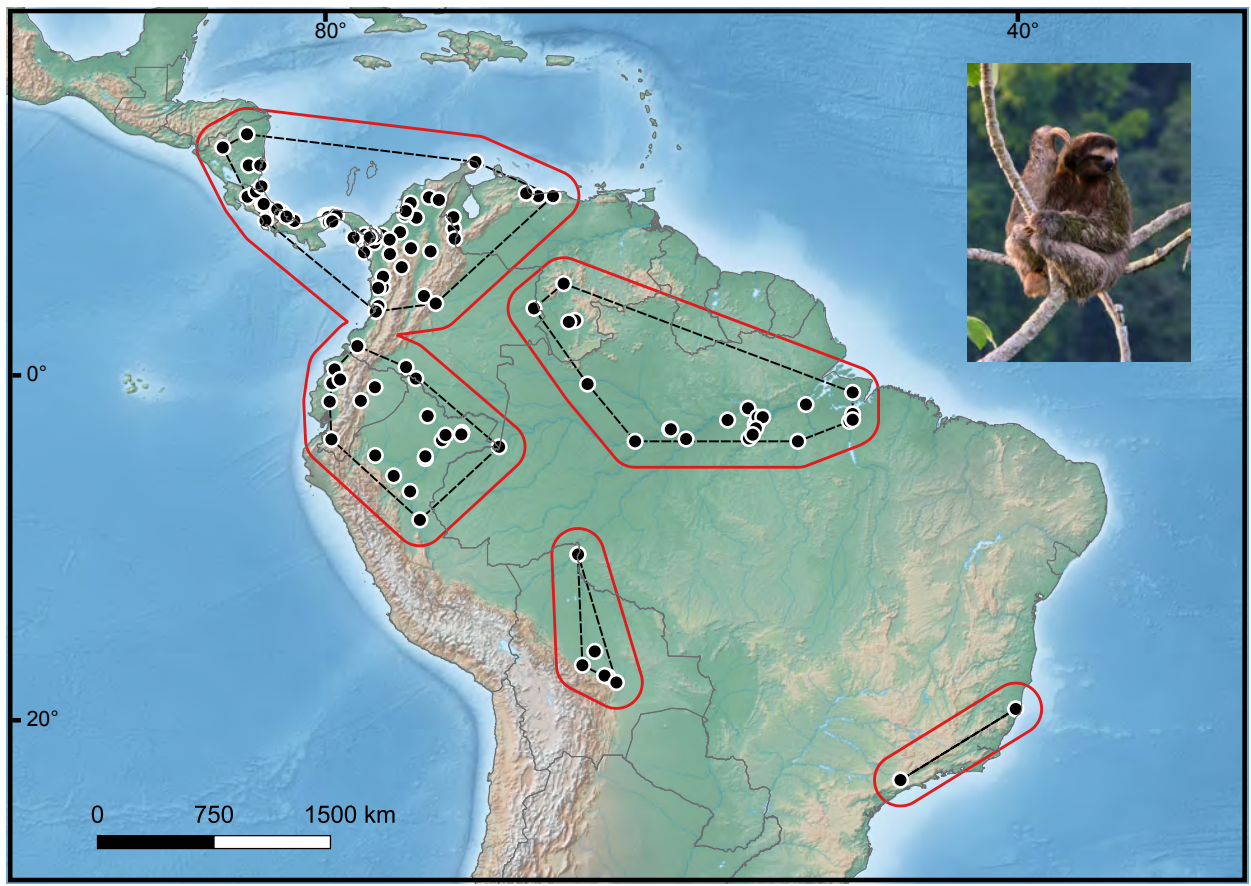

**Figure 1.** Occurrence localities and study area employed to calibrate ecological niche models of Bradypus variegatus. Dots represent occurrences of the species. Dashed lines show minimum convex polygons around major clusters of occurrences. To create the polygons that collectively conformed our study region for model calibration (shown with solid lines) an external buffer (with a distance of 1.5 degrees) was created from each of the minimum convex polygons.

All of the ENM analyses considered 17 bioclimatic obtained from WorldClim 1.4 (Table 1; Hijmans et al. 2005; http://www.worldclim.org) relevant for the period ca. 1960–1990. We calibrated models employing climatic scenarios for that period ("Present"), and models were then projected on this and others climatic scenarios employing multiple Global Circulation Models (Table 2) and an extent defined by the geographic coordinates: 20.25° N–29.75º S and 92.25° W–33.75º W. Study regions for the model calibration were selected following fundamental principles for such purpose (Anderson & Raza, 2010; Barve et al., 2011; Fig. 1; Appendix S1).

**Table 1.** Code and brief description of each environmental variable employed to calibrate ecological niche models. These data were obtained from WorldClim 1.4 (Hijmans et al. 2005; http://www.worldclim.org). A quarter is a period of three months (1/4 of the year).

| Code | Bioclimatic variable |
| --- | --- |
| BIO1 | Annual Mean Temperature |
| BIO2 | Mean Diurnal Range (Mean of monthly (max temp - min temp)) |
| BIO3 | Isothermality (BIO2/BIO7) (* 100) |
| BIO4 | Temperature Seasonality (standard deviation *100) |
| BIO5 | Max Temperature of Warmest Month |
| BIO6 | Min Temperature of Coldest Month |
| BIO7 | Temperature Annual Range (BIO5-BIO6) |
| BIO8 | Mean Temperature of Wettest Quarter |
| BIO9 | Mean Temperature of Driest Quarter |
| BIO10 | Mean Temperature of Warmest Quarter |
| BIO11 | Mean Temperature of Coldest Quarter |
| BIO12 | Annual Precipitation |
| BIO13 | Precipitation of Wettest Month |
| BIO14 | Precipitation of Driest Month |
| BIO15 | Precipitation Seasonality (Coefficient of Variation) |
| BIO16 | Precipitation of Wettest Quarter |
| BIO17 | Precipitation of Driest Quarter |

***Tuning model parameters.—***We employed preliminary models to tune MaxEnt's feature classes (FCs) and regularization multiplier (RM) (Appendix S1). FCs are functions of the original environmental variables. RM is a coefficient that penalizes complexity, and preserves the relative strengths of the regularization parameter beta ($ß$) across feature classes used to produce a model (Anderson & Gonzalez, 2011; Shcheglovitova & Anderson, 2013). We employed the R package "ENMeval" (Muscarella et al., 2014) and MaxEnt (ver. 3.4.1). The following combinations of FCs were considered: "H", "LQH", "LQP", and "LQPH". For each of these FCs, or combinations thereof, we conducted ten analyses, one with each of the RMs in the range 0.5–5.0 with increments of 0.5, with the geographic partition scheme "block" of ENMeval.

**Table 2.** Climatic scenarios employed for calibration and/or projection of Ecological Niche Models. These climatic scenarios were obtained from WorldClim 1 (Hijmans et al. 2005). Time is in either number of years before present (for Last Interglacial, Last Glacial Maximum, Mid Holocene) or as year of Gregorian calendar (for Present, Future). The resolution of corresponding raster datasets is expressed in decimal degrees. GCM: Global Circulation Model; RCP: Representative Concentration Pathway.

| Period name | Time | Resolution | GCM | RCP |
|---|---|---|---|---|
| Last Interglacial (LIG) | ~120–140 ka | 0.5 | CCSM3 | — |
| Last Glacial Maximum (LGM) | ~22 ka | 2.5 | CCSM4 | — |
| | | | MPI-ESM-P | — |
| | | | MIROC-ESM | — |
| Mid Holocene (MH) | ~6 ka | 2.5 | CCSM4 | — |
| | | | MPI-ESM-P | — |
| | | | MIROC-ESM | — |
| Present | ~1960–1990 | 2.5 | — | — |
| Future (2070) | ~2060–2080 | 2.5 | CCSM4 | 2.6, 4.5, 6.0, 8.5 |
| | | | MPI-ESM-LR | 2.6, 4.5, 8.5 |
| | | | MIROC-ESM | 2.6, 4.5, 6.0, 8.5 |

***Optimality criteria and calibration of final models.***—Preliminary models were examined with four optimality criteria to select the "best" parameters for calibration of a final model (one per optimality criterion). Two optimality criteria were based on metrics of model accuracy: omission rates (ORs) and Area Under the Curve of the Receiver Operating Characteristic plot (AUC) (see definitions in Appendix S1). The former were calculated with two threshold rules, the 10-percentile threshold (10P) and the lowest presence threshold (LPT; Pearson et al., 2007); we refer to the ORs obtained with these thresholds by the abbreviations $OR_{10}$ and $OR_{LPT}$, respectively.

Two other optimality criteria were based on the Akaike Information Criterion (AIC; Akaike, 1973). To measure how much each model in the set of candidate models deviates from a model perfectly representing the reality of a studied phenomenon or process without errors (the "truth"), the Kullback-Leibler distance is employed (Burnham & Anderson, 2002). The AIC serves to estimate the expected value of the relative Kullback-Leibler distance without knowing the truth (Hobbs & Hilborn, 2006). We employed the correction of AIC for small sample sizes, AICc, for all analyses and text (Burnham & Anderson, 2004). One of our optimality criteria favors model parameters that yield the lowest AICc score; hereafter we refer to this optimality criterion as "LowAICc".

Each model in a set of compared models has biases associated to its estimated parameters leading to divergence among model predictions—a phenomenon known as model uncertainty (Burnham & Anderson, 2002). In our analyses distinct parameterizations are a consequence of the use of different FCs and RMs. As optimality criterion, AICc often is used to select a single model, the one with the lowest AICc value, discarding all other models. This procedure might be acceptable when one model is clearly the best (e.g., a model whose AICc weight ≥ 0.85) among all models being considered. However, if no single model were clearly superior to the others, then substantial amount of useful predictive power would be wasted if only one model is used to make predictions. The uncertainty of the strength of evidence among models can be measured through AICc weights ($w_i$AICc), which is a convenient normalization of Akaike scores, so that all AICc scores sum to 1 (see Hobbs & Hilborn, 2006).

Akaike weights enable that each selected model could be employed for inference, with each model contributing to a consensus prediction proportionally to its relative weight of evidence for explaining the data. This is known as "AIC model averaging consensus" or simply "AIC model averaging" (AvgAICc) and can be easily implemented in ENM by weighting the prediction (suitability values) of each model using $w_i$AICc. In the context of ENM, AvgAICc can serve as strategy for multi-model prediction. As such, the prediction (suitability values) of each selected model is multiplied by the $w_i$AICc of the corresponding model. See Appendix S1 for a description on how corresponding threshold rules can be applied when this approach is implemented.

Calculating the averaged importance and relative contribution of individual variables is straightforward. All MaxEnt models contain exactly the same predictor (environmental) variables, but differ only in variable transformations (i.e. Feature Classes; Phillips et al., 2006; Phillips & Dudík, 2008; Elith et al., 2011). MaxEnt computes the relative importance and contribution of individual variables on each model, which can be multiplied by the $w_i$AICc of the corresponding model to obtain the averaged importance and contribution of individual variables across models. We provide an R script (Appendix S2) that allows reproducibility of all the analyses conducted for this study.

*Model comparisons.*—The performance of models and predictions resulting from the application of the AvgAICc approach were compared with those of final models selected under other optimality criteria. Model complexity was compared through the number of model parameters. Prediction pattering was compared quantitatively and qualitatively. For the former we calculated the area ($km^2$) predicted suitable for the focal species employing the 10P and LPT thresholds rules as well as with the sensitivity-specificity equality threshold rule (also known as "equal training sensitivity and specificity threshold", ETSS; see Pearson et al., 2004; Peterson et al., 2011). For qualitative assessments, we examined pairwise comparisons of prediction strengths projected on geographic space. This procedure enabled detection of both: (1) regions where only one model inferred suitable conditions for the focal species; and (2) regions where the two compared models identified suitable conditions. Finally, we compared the relative contributions of each environmental variable to the models.

## RESULTS

*The AvgAICc strategy.*—The application of the AvgAICc approach yielded a prediction assembled with 14 models ($\sum w_i$AICc = 0.992). Each of these models resulted from an analysis based on all filtered occurrences and carried out with different combinations of FCs and/or RMs (Table 3). Each of the four top-ranked models had a $w_i$AICc ≥ ca. 0.057, and the sum of their $w_i$AICc was ca. 0.77. Three combinations of FCs were among these top-four ranked models, as follows: Linear Quadratic Hinge (LQH); Linear Quadratic Product (LQP); and Linear Quadratic Product Hinge (LQPH). These four models had RMs of either 3, 3.5, or 4, and their numbers of parameters ranged from 9 to 13 (Table 3). Among the remaining models (occupying lower ranks of the AICc hierarchy) the highest numbers of parameters were 15 and 17, but the corresponding models had quite marginal AICc weights (Fig. 2; Table 3). The AvgAICc criterion performed well—with regard to model complexity, omission rates, AUC—compared to models favored by other optimality criteria in the model tuning analyses. Detailed comparisons between the AvgAICc criterion and other criteria are presented below.

**Table 3.** Models selected with different optimality criteria for *Bradypus variegatus* and their corresponding complexity and performance. Optimality criteria: $OR_{10}$, omission rate calculated with the 10-percentile threshold (10P); $OR_{LPT}$, omission rate calculated with the lowest presence threshold (LTP, Pearson et al., 2007); AUC, area under the curve of the Receiver Operating Characteristic; AICc, score of the Akaike Information Criteria; AICc average, average model whose consensus prediction was pondered with the weight of preliminary models calculated through the parameter tuning analyses. Variables: FC and RM, feature classes and regularization multiplier employed by MaxEnt, respectively; wAICc, AICc weight; NP, number of model parameters; R, rank in the AICc hierarchy (sorted from best to worst). NA values are present when the number of parameters is larger than the number of occurrence localities.

| Optimality criteria | FC | RM | AICc | $w_iAICc$ | NP | Rank | $OR_{10}$ | $OR_{LPT}$ | AUC |
|---|---|---|---|---|---|---|---|---|---|
| AICc 1 | LQH | 3 | 2626.56 | 0.46 | 10 | 1 | 0.259 | 0.148 | 0.628 |
| AICc 2 | LQH | 3.5 | 2628.28 | 0.19 | 9 | 2 | 0.241 | 0.139 | 0.621 |
| AICc 3 | LQP | 4 | 2630.51 | 0.06 | 12 | 3 | 0.222 | 0.111 | 0.620 |
| AICc 4 | LQPH | 4 | 2630.73 | 0.06 | 13 | 4 | 0.222 | 0.139 | 0.625 |
| AICc 5 | LQH | 5 | 2631.15 | 0.05 | 7 | 5 | 0.231 | 0.120 | 0.615 |
| AICc 6 | LQH | 4.5 | 2631.36 | 0.04 | 8 | 6 | 0.231 | 0.130 | 0.618 |
| AICc 7 | LQH | 4 | 2631.45 | 0.04 | 9 | 7 | 0.241 | 0.130 | 0.619 |
| AICc 8 | LQPH | 3.5 | 2632.22 | 0.03 | 15 | 8 | 0.231 | 0.139 | 0.629 |
| AICc 9 | LQP | 3.5 | 2632.47 | 0.02 | 14 | 9 | 0.231 | 0.120 | 0.624 |
| AICc 10 | LQP | 4.5 | 2633.88 | 0.01 | 12 | 10 | 0.222 | 0.093 | 0.617 |
| AICc 11 | LQPH | 3 | 2633.88 | 0.01 | 17 | 11 | 0.259 | 0.139 | 0.629 |
| AICc 12 | LQPH | 5 | 2634.82 | 0.01 | 11 | 12 | 0.213 | 0.120 | 0.615 |
| AICc 13 | LQP | 5 | 2634.98 | 0.01 | 11 | 13 | 0.222 | 0.083 | 0.614 |
| AICc 14 | LQPH | 4.5 | 2635.24 | 0.01 | 13 | 14 | 0.213 | 0.130 | 0.620 |
| LowAICc | LQH | 3 | 2626.56 | 0.46 | 10 | 1 | 0.259 | 0.148 | 0.628 |
| $OR_{LPT}$ (2°: AUC) | LQP | 5 | 2634.98 | 0.01 | 11 | 13 | 0.222 | 0.083 | 0.614 |
| $OR_{10}$ (2°: AUC) | LQPH | 4.5 | 2635.24 | 0.01 | 13 | 14 | 0.213 | 0.130 | 0.620 |
| AUC (2°: $OR_{LPT}$) | LQPH | 1 | 2961.55 | 0.00 | 72 | 37 | 0.259 | 0.130 | 0.686 |
| AUC (2°: $OR_{10}$) | LQPH | 1 | 2961.55 | 0.00 | 72 | 37 | 0.259 | 0.130 | 0.686 |

***Model complexity.***—The LowAICc criterion favored a model that has fewer parameters than models selected by any of other optimality criteria (Fig. 2; Table 3); however, the AvgAICc criterion selected models that in average had only slightly higher number of parameters than that selected by the LowAICc criterion. Moreover, as noted earlier, almost all of the models considered under the AvgAICc approach that had higher numbers of parameters than the model selected by the LowAICc criterion had very small AICc weights (Fig. 2; Table 3). Optimality criteria based on metrics of accuracy yielded preliminary models whose number of parameters varied from being similar to those selected by AICc-based criteria to some with substantially more parameters. The former corresponded to the omission rates-based criteria and the latter to the AUC-based criteria (which had 72 parameters)! (Fig. 2, Table 3).

Among the models selected by the AvgAICc optimality criterion were models that included the only two combinations of FCs that were favored by the other optimality criteria (Table 3). By contrast, among the models selected under the AvgAICc criterion was a combination of FCs, i.e., Linear Quadratic Hinge, which was not selected by any of the accuracy metric-based optimality criteria. This was the combination of FCs also of the model selected by LowAICc criterion, but the LowAICc missed a series of combinations of FCs and RMs that were included in 13 of the 14 models selected by the AvgAICc criterion and that collectively account for a cumulative $w_i$AICc of ca. 0.54. With respect to RMs, the two AICc-based criteria and the two omission rate-based criteria selected models with somewhat similar values, although the models with larger $w_i$AICc in the AvgAICc criterion had slightly lower RMs (Table 3). The two optimality criteria based on AUC selected models with RMs equal 1 (Table 2), which coincides with the value of RM used by MaxEnt when default settings are implemented.

***Model evaluation.***—All optimality criteria selected preliminary models that performed better than a random model, as their AUC were above 0.5 (Fig. 2; Table 3). AUC values ranged from ca. 0.614 to ca. 0.686. The five top-ranked models selected under the AvgAICc criterion—which together account for an accumulative AIC weight ≥ 0.816—had AUC values above ca. 0.615.

All of the compared optimality criteria selected models that had omission rates (ORs) higher than the expected theoretical ORs for corresponding threshold rules (Fig. 2; Table 3). For $OR_{10}$ values ranged from ca. 0.213 to ca. 0.259, whereas for $OR_{LPT}$ values ranged from ca. 0.083 to ca. 0.148. The top five-ranked models selected under the AvgAICc had average ORs of ca. 0.235 and 0.131 for $OR_{10}$ and $OR_{LPT}$, respectively (Table 3).

***Contribution of environmental variables.***—Although two environmental variables had the largest contributions to most models favored by different optimality criteria, some discrepancies exist (Fig. 3; Appendix S3, Table S3.1). The variable that had the largest contribution under the AvgAICc criterion was bio13 (the precipitation of wettest month; ca. 35%) followed by bio2 (the mean diurnal range of temperature; ca. 33%; Fig. 3). These same variables were the two that contributed the most to the model favored by the LowAIC criterion, but in this case bio2 contributed (ca. 37%) more than bio13 (ca. 33%; Fig. 3). In the 14 models that summed a $_w$AIC

> 0.99 in the AvgAICc criterion—as well as in the models favored by criteria based on metrics of accuracy—all 17 variables had some contribution to the prediction, whereas in the model favored by the Lowest AIC criterion five variables had zero contribution. In the models preferred under the accuracy metric-based criteria a few environmental variables (e.g., bio3, bio12) had moderate contributions, but these variables had only truly marginal contributions in models favored by the AIC-based criteria (Fig. 3).

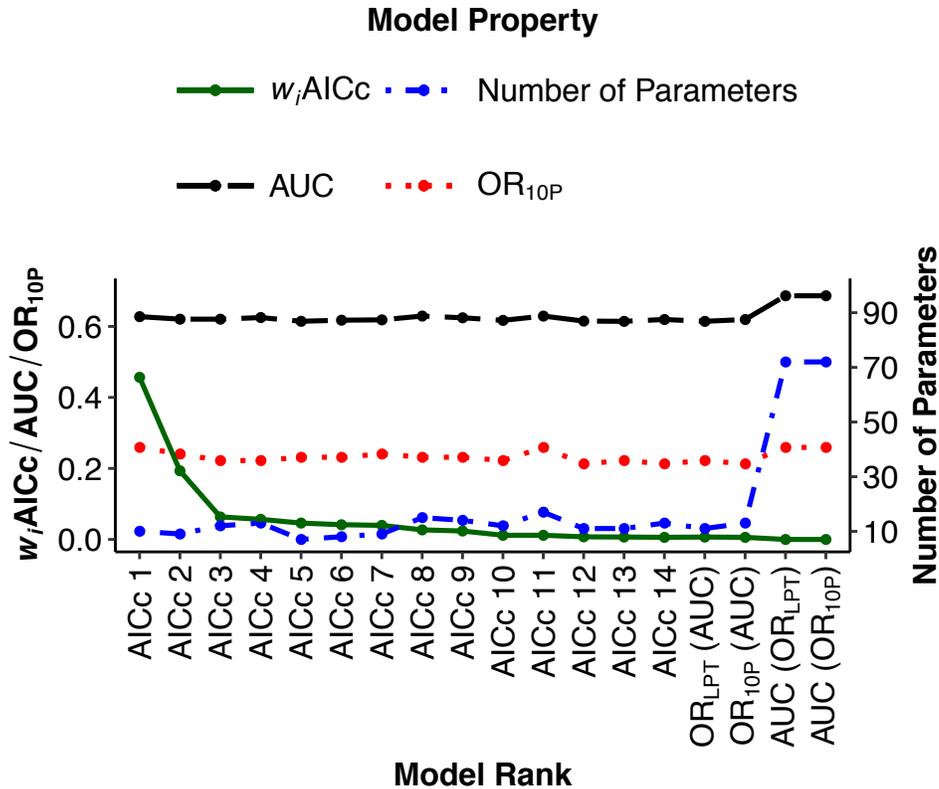

**Figure 2.** Properties of models whose parameters were tuned-up with different optimality criteria. Displayed properties are: Akaike weights ($w_i$AICc), Area Under the Curve of the Receiver Operating Characteristic plot (AUC), omission rate calculated based on the 10-percentile threshold ($OR_{10p}$), and number of parameters of each model. The optimality criteria employed to tune-up the regularization multipliers and combinations of feature classes of MaxEnt were: $OR_{LPT}$ *(AUC)*, omission rate calculated based on the lowest training presence threshold (using AUC as a secondary criteria, if needed); $OR_{10P}$ *(AUC)*, omission rate calculated based on the 10-percentile threshold (and using AUC as a secondary criteria, if needed); *AUC* $(OR_{LPT})$, the Area Under the Curve of the Receiver Operating Characteristic plot (using $OR_{LPT}$ as a secondary criteria, if needed); *AUC* $(OR_{10P})$, the Area Under the Curve of the Receiver

Operating Characteristic plot (using $OR_{10P}$ as a secondary criteria, if needed), lowest AICc (LowAICc), and the 14 AICc top-ranked models used for averaging the final prediction (AvgAICc).

***Prediction pattering.***—The amount of area predicted as holding suitable environmental conditions for the focal species varied among models selected with different optimality criteria (; Appendix S3, Table S3.2). As expected, large differences were observed in comparisons between models whose parameters were tuned-up with AIC-based optimality criteria and those tuned-up with accuracy-metrics-based criteria. However, important differences were also observed between predictions of both AIC-based criteria, for example, when the models were projected onto climatic scenarios of the mid-Holocene and the Present (Fig. 4; Appendix S3, Table S3.2). In general, models favored by the omission rate-based criteria were the ones that estimated the largest suitable areas, whereas models selected by AIC-based criteria estimated considerably smaller suitable areas.

Models selected under different optimality criteria also showed important differences with respect to the geographic location of predicted suitable climatic conditions for the focal species (Figures in Gutiérrez & Heming, 2018). Particularly interesting in the context of this article, in which the AvgAICc-based optimality criterion and predictive strategy is introduced, is the fact that it predicted suitable conditions in geographic areas where the LowAICc criterion did not (Fig. 5). This is caused by differences in predictions by models favored by different optimality criteria (see *Model prediction uncertainty* section below). The difference between predictions is more patent on projections of both of these models onto the climatic scenarios of the: (1) mid-Holocene, mainly in northwestern Brazil; (2) Present, in southeastern Nicaragua, southern Panama, the southern Llanos of central Venezuela, and on central and southern Brazil; year 2070, in the southern Llanos of central Venezuela and on central and southern Brazil (Fig. 5). The model favored by the AvgAICc criterion predicted suitable conditions in these areas for the just mentioned climatic scenarios, whereas the model that was favored by the LowAICc criterion did not. Comparisons between these two models (i.e., AICc-based) and those tuned-up with accuracy-based criteria revealed even more substantial differences (Figures in Gutiérrez & Heming, 2018).

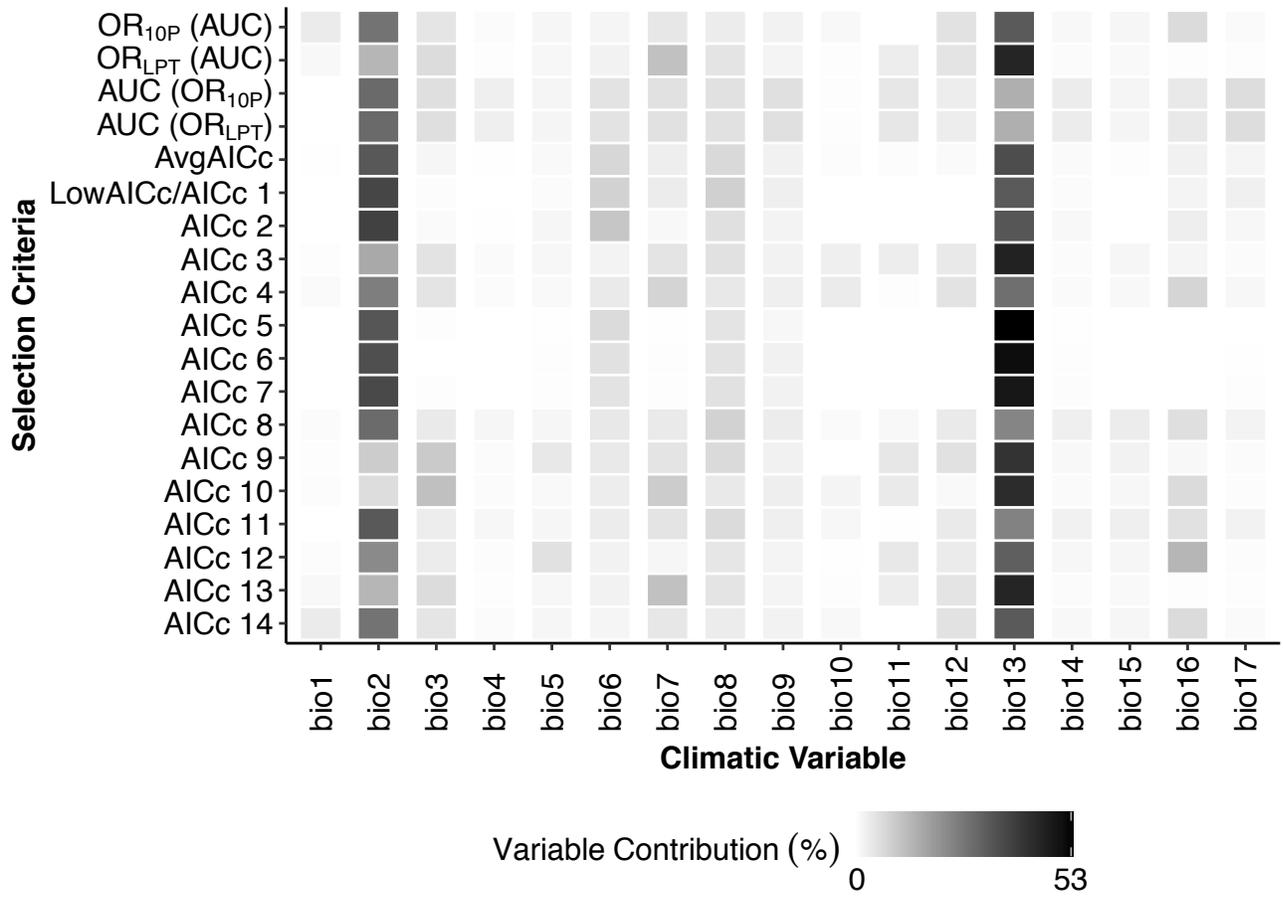

**Figure 3.** Relative contribution of environmental variables to models whose parameters were tuned-up with different optimality criteria. The optimality criteria employed to tune-up the regularization multipliers and combinations of feature classes of MaxEnt were: $OR_{LPT}$ *(AUC)*, omission rate calculated based on the lowest training presence threshold (using AUC as a secondary criteria, if needed); $OR_{10P}$ *(AUC)*, omission rate calculated based on the 10-percentile threshold (and using AUC as a secondary criteria, if needed); *AUC ($OR_{LPT}$)*, the Area Under the Curve of the Receiver Operating Characteristic plot (using $OR_{LPT}$ as a secondary criteria, if needed); *AUC ($OR_{10P}$)*, the Area Under the Curve of the Receiver Operating Characteristic plot (using $OR_{10P}$ as a secondary criteria, if needed), lowest AICc (LowAICc), and the 14 AICc top-ranked models used for averaging the final prediction (AvgAICc).

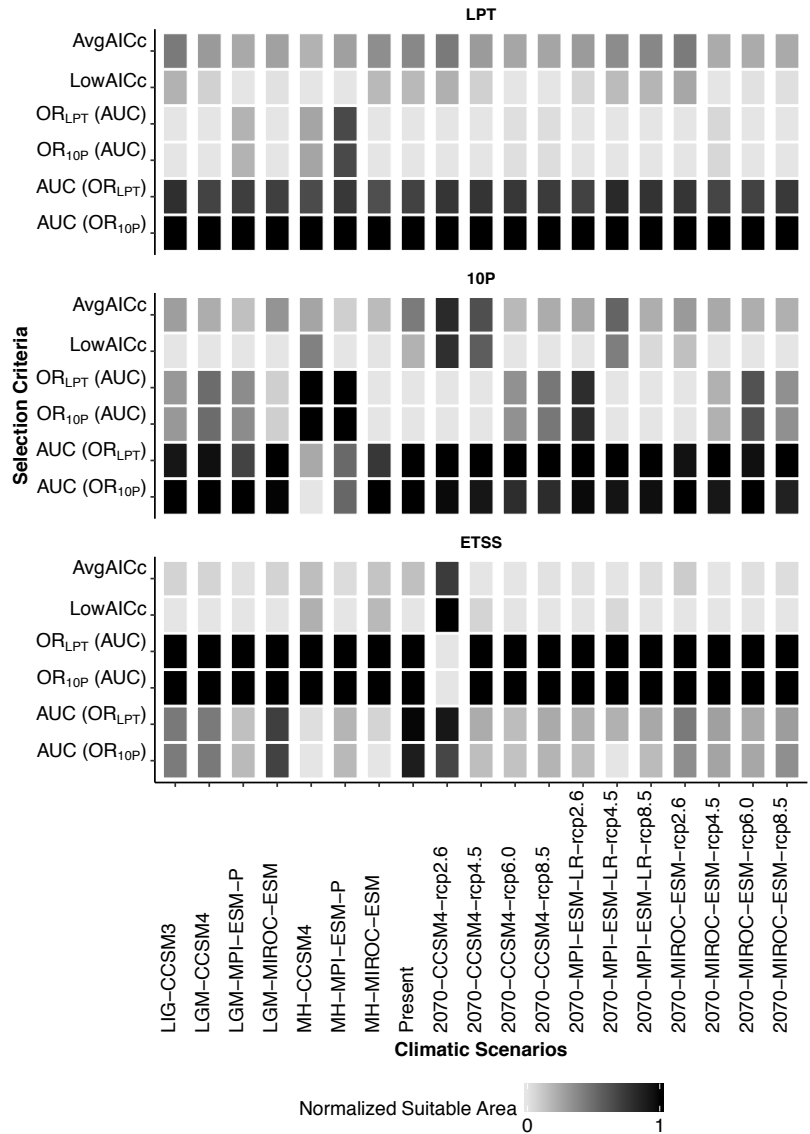

**Figure 4.** Amount of habitat predicted suitable for *Bradypus variegatus* by models whose parameters were tuned-up with different optimality criteria. The amount of predicted suitable area is expressed in million of km$^2$ and was calculated after applying a 10-percentile threshold to covert continue model predictions into binary ones. The models were projected onto climatic conditions of both the mid-Holocene (using three global circulations models) and the Present (see Methods for further details). The optimality criteria employed to tune-up the regularization multipliers and combinations of feature classes of MaxEnt were as follows: *AvgAICc*, the AIC model averaging optimality criteria and prediction strategy; *LowAICc*, the model with the lowest AIC scores; *OR$_{LPT}$ (AUC)*, omission rate calculated based on the lowest training presence threshold (using AUC as a secondary criteria, if needed); *OR$_{10P}$ (AUC)*, omission rate calculated

based on the 10-percentile threshold (and using AUC as a secondary criteria, if needed); *AUC (OR$_{LPT}$)*, the Area Under the Curve of the Receiver Operating Characteristic plot (using *OR$_{LPT}$* as a secondary criteria, if needed); *AUC (OR$_{10P}$)*, the Area Under the Curve of the Receiver Operating Characteristic plot (using *OR$_{10P}$* as a secondary criteria, if needed)

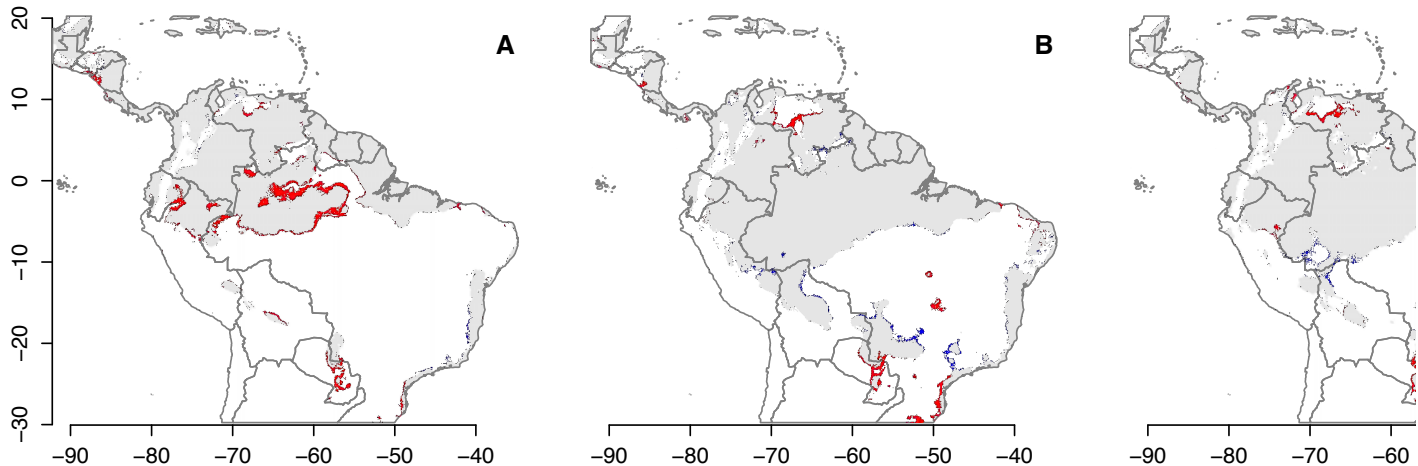

**Figure 5.** Comparison of geographic projections of models tuned-up with the AvgAICc-based and LowAICc optimality criteria. The comparisons were made with models projected onto climatic conditions of (A) the mid-Holocene (ca. 6 ka; derived from the MPI-ESM-P general circulation model); (B) the Present (average for period comprising the years 1960–1990); (C) the year 2070 (average for period comprising the years 2061–2080; derived from the CCSM4 general circulation model and 8.5 representative concentration pathway). Projections are binary representations (i.e., predicted presence/absence of suitable climatic conditions for *Bradypus variegatus*) after applying the 10-percentile threshold. In red are shown areas where the model favored by the AvgAICc criterion predicted suitable conditions and the model favored by the LowAICc did not. In blue are shown areas where the model favored by the LowAICc criterion predicted suitable conditions and the model favored by the AvgAICc did not. Areas in grey were predicted suitable by both models.

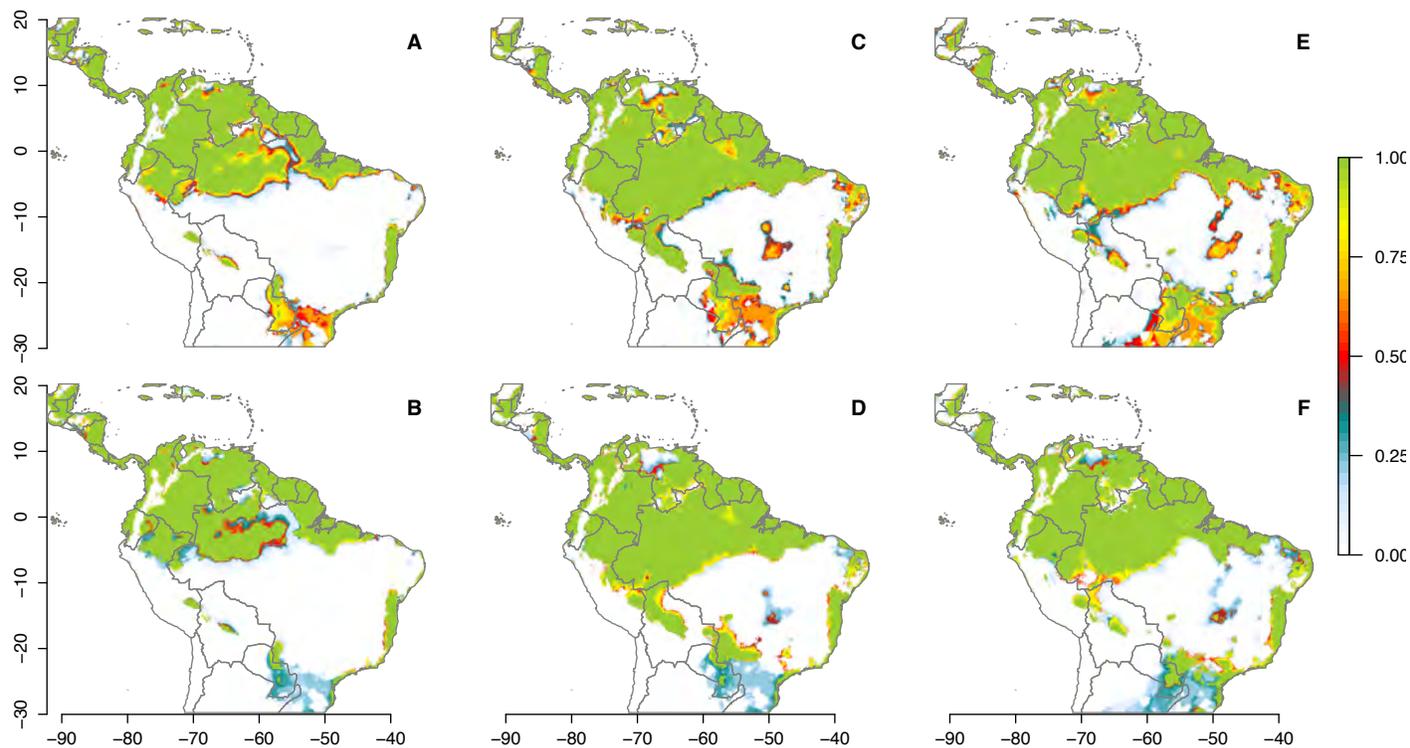

**Figure 6.** Uncertainties among geographic projections of 14 selected models used in AvgAICc final prediction. Values correspond to the proportion of models predicting each site as suitable (i.e., predicted presence/absence of suitable climatic conditions) for the focal species (*Bradypus variegatus*) after applying the 10-percentile threshold. Top row (A, C, E) correspond to the raw proportion of models that predicted the area as suitable. Bottom row (B, D, F) correspond to AICc weighted proportion of models that predicted the area as suitable. The comparisons were made with models projected onto climatic conditions of (A, B) the mid-Holocene (ca. 6 ka; derived from the MPI-ESM-P general circulation model); (C, D) the Present (average for period comprising the years 1960–1990); (E, F) the year 2070 (average for period comprising the years 2061–2080; derived from the CCSM4 general circulation model and 8.5 representative concentration pathway). Uncertainty is high when the proportion of models that predict the any area as suitable/unsuitable approximates 0.5 and decreases when this proportion approximates both 0 and 1 (i.e. there is more consensus among model predictions, both for species absence and presence).

***Model prediction uncertainty.—***Most of geographic locations that were predicted as suitable/unsuitable for the focal species were congruent (Fig. 6) among the 14 models ($\sum w_i$AICc = 0.992) used to build the final AvgAICc prediction. However, in a relatively large geographic extent the suitability was not congruent among these models (i.e., model uncertainty; Fig. 6A, 6C, and 6E). When model uncertainty was weighted using $w_i$AICc ("weighted uncertainty"), then the geographic extent of areas with high uncertainty was drastically reduced (Fig. 6B, 6D, and 6F). Most of the areas with high "weighted uncertainty" were the same of those included in the geographic distribution of the final AvgAICc prediction.

## DISCUSSION

We herein introduce a model average approach based on the Akaike Information Criterion, i.e., the AvgAICc approach, to serve both as an optimality criterion in analyses to tune-up parameters of ecological niche models and as a multi-model prediction strategy. The AvgAICc approach yields results that differ from those of alternative approaches previously used in a number of studies (e.g., Anderson & Gonzalez, 2011; Shcheglovitova & Anderson, 2013; Gutiérrez et al., 2014; Muscarella et al., 2014; Moreno-Amat et al., 2015; Soley–Guardia et al., 2016). The differences are in all of the aspects considered in this study, all of which of importance for all applications of ENM. These aspects are model complexity, selection of environmental variables, and amount and geographic location of areas predicted as suitable for the modeled species.

The two AICc-based approaches consistently yielded similar, but not identical, results. In the AvgAICc approach the prediction comprised 14 models ($\sum w_i$AICc = 0.9918), including the same (single) model favored by the LowAICc criterion plus 13 others. The model selected by the LowAICc criterion accounted for 0.46 of the total $w_i$AICc (i.e., 1.00), while the other 13 models collectively accounted for more than half (ca. 0.54) of the total AICc weights, thus explaining the differences found between the predictions resulting from these approaches. These differences are not trivial, as they might heavily impact the effectiveness of ENM analyses applied to many kinds of studies. This is especially due to the fact that AvgAICc approach seems to be able to detect suitable conditions in geographic areas where the LowAICc approach fails to do so. This happens because when taking into account a single model (i.e., LowAICc) to make predictions, there is valuable information contained on other models that is ignored. When alternative models (e.g., with differing parameters) are closely ranked to the best model, there is uncertainty about which model better represents the observed data (Fig. 6). Predictions based on a single model may vary substantially depending on model parameterization (Burnham & Anderson, 2002), and in practical terms this may lead to both overlooking suitable areas for the focal species and incorrectly predicting as suitable areas do not harbor suitable conditions for the modeled species—i.e. increased omission and commission errors, respectively. When taking into account the uncertainty associated to each model by averaging the contribution of each individual model (i.e., AvgAICc) it is expected to obtain more robust predictions that are relatively more precise and less biased than those based on single models (Fig. 6, e.g., LowAICc; Burnham & Anderson, 2002). This approach, which takes into account information that would be otherwise ignored, is expected to provide more reliable predictions and reduce uncertainties associated to model transferability.

The application of the AvgAICc approach might increase the effectiveness of ENM applications. To name a few examples, the AvgAICc approach might be able to identify connections among patches of areas harboring suitable environmental conditions for species whose habitat have been fragmented. Similarly, it might be able to reveal the existence of isolated and/or unexplored regions harboring suitable conditions for focal species, thus promoting fieldwork leading to the discovery of likely taxonomically and ecologically interesting populations (e.g., Raxworthy et al., 2003; Giacomin et al., 2014), or to collecting samples of rare species (e.g., Oleas et al., 2014). These discoveries would be hampered if the parameter-tuning analyses that precede the final model calibration processes were carried out using optimality criteria that favor models unable to spot some of the regions harboring suitable conditions for the focal species—as is the case of the LowAICc approach, which in some of our analyses failed to predict suitable conditions in areas where the AvgAICc approach did not.

Model averages based on AIC weights to assess the relative importance of predictors have been recently criticized on three grounds, as follows (Cade (2015): (1) multicollinearity among predictor variables might cause that neither the parameters of the models nor their estimates have common scales; (2) the associated sums of AIC weights are a measure of relative importance of models, with little information about contributions by individual predictors; (3) sometimes the model-averaged regression coefficients for predictor variables are incorrectly used to make model-averaged predictions of the response variable when the models are not linear in the parameters. Cade's criticisms have not been universally accepted (e.g., Walker, 2017), but regardless, none of those alleged issues apply to the AvgAICc approach we herein introduce into ENM. This is because, (1 and 3) when creating the averaged prediction we did not average coefficients of each model, but averaged the final predictions of each model; (2) unlike model selection in other algorithms (e.g. GLM, and those for occupancy modeling)—where models are built using different predictor variables and the variable relative importance is the weight of the model where the variable is present (Burnham & Anderson, 2002; Cade, 2015)—in this study, the averaged importance and contribution of individual variables across models does involve their individual values within models (see Methods).

By introducing the AvgAICc approach in ENM we hope to promote further research that should characterize the pros and cons of this approach relative to others as both optimality criterion and multi-model prediction strategy. Our analyses demonstrate that this approach yields substantially different results that alternative ones in aspects of importance for common applications of ENM. Given the theoretically advantages of the AvgAICc approach—which have led to its wide use in a plethora of fields in science, and particularly in ecology and evolution (Sillett & Holmes, 2002; Posada & Buckley, 2004; Brodziak & Legault, 2005; Diniz-Filho et al., 2008)—we expect that the use of the AvgAICc procedure will become common and useful in ENM, particularly in applications of ENM that require transferences of models to climatic scenarios across space, time, or both.

# REFERENCES

Akaike, H. (1973) Information theory and an extension of the maximum likelihood principle. *Second International Symposium on Information Theory*, 267–281.

Anderson, R.P. (2012) Harnessing the world's biodiversity data: promise and peril in ecological niche modeling of species distributions. *Annals of the New York Academy of Sciences*, **1260**, 66–80.

Anderson, R.P. & Gonzalez, I. (2011) Species-specific tuning increases robustness to sampling bias in models of species distributions: An implementation with Maxent. *Ecological Modelling*, **222**, 2796–2811.

Anderson, R.P. & Raza, A. (2010) The effect of the extent of the study region on GIS models of species geographic distributions and estimates of niche evolution:

preliminary tests with montane rodents (genus *Nephelomys*) in Venezuela. *Journal of Biogeography*, **37**, 1378–1393.

Araújo, M.B. & New, M. (2007) Ensemble forecasting of species distributions. *Trends in Ecology & Evolution*, **22**, 42–47.

Barve, N., Barve, V., Jiménez-Valverde, A., Lira-Noriega, A., Maher, S.P., Peterson, A.T., Soberón, J., & Villalobos, F. (2011) The crucial role of the accessible area in ecological niche modeling and species distribution modeling. *Ecological Modelling*, **222**, 1810–1819.

Boria, R.A., Olson, L.E., Goodman, S.M., & Anderson, R.P. (2017) A single-algorithm ensemble approach to estimating suitability and uncertainty: cross-time projections for four Malagasy tenrecs. *Diversity and Distributions*, **23**, 196–208.

Brodziak, J. & Legault, C.M. (2005) Model averaging to estimate rebuilding targets for overfished stocks. *Canadian Journal of Fisheries and Aquatic Sciences*, **62**, 544–562.

Buisson, L., Thuiller, W., Casajus, N., Lek, S., & Grenouillet, G. (2010) Uncertainty in ensemble forecasting of species distribution. *Global Change Biology*, **16**, 1145–1157.

Burnham, K.P. & Anderson, D.R. (2002) *Model Selection and Multi-Model Inference: A Practical Information-Theoretic Approach.* Springer, New York, New York, USA.

Burnham, K.P. & Anderson, D.R. (2004) Multimodel Inference Understanding AIC and BIC in Model Selection. *Sociological Methods & Research*, **33**, 261–304.

Cade, B.S. (2015) Model averaging and muddled multimodel inferences. *Ecology*, **96**, 2370–2382.

Diniz-Filho, J.A.F., Bini, L.M., Pinto, M.P., Terribile, L.C., de Oliveira, G., Vieira, C.M., Blamires, D., de Souza Barreto, B., Carvalho, P., Rangel, T.F.L.V.B., Tôrres,

N.M., & Bastos, R.P. (2008) Conservation planning: a macroecological approach using the endemic terrestrial vertebrates of the Brazilian Cerrado. *Oryx*, **42**, 567.

Elith, J. & Leathwick, J.R. (2009) Species distribution models: ecological explanation and prediction across space and time. *Annual Review of Ecology, Evolution, and Systematics*, **40**, 677–697.

Elith, J., Phillips, S.J., Hastie, T., Dudík, M., Chee, Y.E., & Yates, C.J. (2011) A statistical explanation of MaxEnt for ecologists. *Diversity and Distributions*, **17**, 43–57.

Galante, P.J., Alade, B., Muscarella, R., Jansa, S.A., Goodman, S.M., & Anderson, R.P. (2017) The challenge of modeling niches and distributions for data-poor species: a comprehensive approach to model complexity. *Ecography*, **41**, 726–736.

Gardner, A.L. (2008) Family Bradypodidae Gray, 1821. *Mammals of South America: Marsupials, Xenarthrans, Shrews, and Bats* (ed. by A.L. Gardner), pp. 158–164. University of Chicago Press, Chicago, Illinois, U.S.A.

Giacomin, L.L., Kamino, L.H.Y., & Stehmann, J.R. (2014) Speeding up the discovery of unknown plants: a case study of Solanum (Solanaceae) endemics from the Brazilian Atlantic Forest. *Boletim do Museu de Biologia Mello Leitão. Nova Série*, **35**, 121–135.

Guisan, A., & Thuiller, W. (2005). Predicting species distribution: offering more than simple habitat models. Ecology Letters, 8, 993–1009.

Gutiérrez, E.E., Boria, R.A., & Anderson, R.P. (2014) Can biotic interactions cause allopatry? Niche models, competition, and distributions of South American mouse opossums. *Ecography*, **37**, 741–753.

Gutiérrez, E.E. & Heming, N.M. (2018) Partial supporting material for article: 'Introducing AIC model averaging in ecological niche modeling: a single-algorithm multi-model strategy to account for uncertainty in suitability predictions" This material will be archived in the public database Zenodo and will receive a doi number. [it is temporarily available for reviewers at https://www.dropbox.com/sh/1wroxf8l1ud2cga/AADxNl--jpXorJF2G6xxh9-7a?dl=0 ]

Halvorsen, R., Mazzoni, S., Bryn, A., & Bakkestuen, V. (2015) Opportunities for improved distribution modelling practice via a strict maximum likelihood interpretation of MaxEnt. *Ecography*, **38**, 172–183.

Hayssen, V. (2010) *Bradypus variegatus* (Pilosa: Bradypodidae). *Mammalian Species*, 19–32.

Hijmans, R.J., Phillips, S., & Elith, J.L. and J. (2017) *dismo: Species Distribution Modeling.*

Hobbs, N.T. & Hilborn, R. (2006) Alternatives To Statistical Hypothesis Testing In Ecology: A Guide To Self Teaching. *Ecological Applications*, **16**, 5–19.

Moreno-Amat, E., Mateo, R.G., Nieto-Lugilde, D., Morueta-Holme, N., Svenning, J.-C., & García-Amorena, I. (2015) Impact of model complexity on cross-temporal transferability in Maxent species distribution models: An assessment using paleobotanical data. *Ecological Modelling*, **312**, 308–317.

Moussalli, A., Moritz, C., Williams, S.E., & Carnaval, A.C. (2009) Variable responses of skinks to a common history of rainforest fluctuation: concordance between phylogeography and palaeo-distribution models. *Molecular Ecology*, **18**, 483–499.

Muscarella, R., Galante, P.J., Soley-Guardia, M., Boria, R.A., Kass, J.M., Uriarte, M., & Anderson, R.P. (2014) ENMeval: An R package for conducting spatially independent evaluations and estimating optimal model complexity for Maxent ecological niche models. *Methods in Ecology and Evolution*, **5**, 1198–1205.

Oleas, N.H., Meerow, A.W., Feeley, K.J., Gebelein, J., & Francisco-Ortega, J. (2014) Using species distribution models as a tool to discover new populations of

Phaedranassa brevifolia Meerow, 1987 (Liliopsida: Amaryllidaceae) in Northern Ecuador. *Check List*, **10**, 689–691.

Pearson, R.G., Dawson, T.P., & Liu, C. (2004) Modelling species distributions in Britain: a hierarchical integration of climate and land-cover data. *Ecography*, **27**, 285–298.

Pearson, R.G., Raxworthy, C.J., Nakamura, M., & Townsend Peterson, A. (2007) Predicting species distributions from small numbers of occurrence records: a test case using cryptic geckos in Madagascar. *Journal of Biogeography*, **34**, 102–117.

Peterson, A.T., Papeş, M., & Eaton, M. (2007) Transferability and model evaluation in ecological niche modeling: a comparison of GARP and Maxent. *Ecography*, **30**, 550–560.

Peterson, A.T. & Soberón, J. (2012) Species distribution modeling and ecological niche modeling: getting the concepts right. *Natureza & Conservação*, **10**, 102–107.

Peterson, A.T., Soberón, J., Pearson, R.G., Anderson, R.P., Martínez-Meyer, E., Nakamura, M., & Araújo, M.B. (2011) *Ecological Niches and Geographic Distributions.* Princeton University Press, Princeton, N.J.

Phillips, S.J., Anderson, R.P., & Schapire, R.E. (2006) Maximum entropy modeling of species geographic distributions. *Ecological Modelling*, **190**, 231–259.

Phillips, S.J. & Dudík, M. (2008) Modeling of species distributions with Maxent: new extensions and a comprehensive evaluation. *Ecography*, **31**, 161–175.

Posada, D. & Buckley, T.R. (2004) Model selection and model averaging in phylogenetics: advantages of akaike information criterion and bayesian approaches over likelihood ratio tests. *Systematic Biology*, **53**, 793–808.

Radosavljevic, A. & Anderson, R.P. (2014) Making better Maxent models of species distributions: complexity, overfitting and evaluation. *Journal of Biogeography*, **41**, 629–643.

Randin, C.F., Dirnböck, T., Dullinger, S., Zimmermann, N.E., Zappa, M., & Guisan, A. (2006) Are niche-based species distribution models transferable in space? *Journal of Biogeography*, **33**, 1689–1703.

Raxworthy, C.J., Martinez-Meyer, E., Horning, N., Nussbaum, R.A., Schneider, G.E., Ortega-Huerta, M.A., & Peterson, A.T. (2003) Predicting distributions of known and unknown reptile species in Madagascar. *Nature*, **426**, 837.

Shcheglovitova, M. & Anderson, R.P. (2013) Estimating optimal complexity for ecological niche models: A jackknife approach for species with small sample sizes. *Ecological Modelling*, **269**, 9–17.

Sillett, T.S. & Holmes, R.T. (2002) Variation in survivorship of a migratory songbird throughout its annual cycle. *Journal of Animal Ecology*, **71**, 296–308.

Soberón, J. (2007) Grinnellian and Eltonian niches and geographic distributions of species. *Ecology Letters*, **10**, 1115–1123.

Soberón, J., Osorio-Olvera, L., & Peterson, T. (2017) Diferencias conceptuales entre modelación de nichos y modelación de áreas de distribución. *Revista Mexicana de Biodiversidad*, **88**, 437–441.

Soley‑Guardia, M., Gutiérrez, E.E., Thomas, D.M., Ochoa‑G, J., Aguilera, M., & Anderson, R.P. (2016) Are we overestimating the niche? Removing marginal localities helps ecological niche models detect environmental barriers. *Ecology and Evolution*, **6**, 1267–1279.

Walker, J.A. (2017) A Defense Of Model Averaging. *bioRxiv*, 133785.

Warren, D.L. & Seifert, S.N. (2011) Ecological niche modeling in Maxent: the importance of model complexity and the performance of model selection criteria.

*Ecological Applications: A Publication of the Ecological Society of America*, **21**, 335–342.

Warren, D.L., Wright, A.N., Seifert, S.N., & Shaffer, H.B. (2014) Incorporating model complexity and spatial sampling bias into ecological niche models of climate change risks faced by 90 California vertebrate species of concern. *Diversity and Distributions*, **20**, 334–343.

Zhang, L., Liu, S., Sun, P., Wang, T., Wang, G., Zhang, X., & Wang, L. (2015) Consensus Forecasting of Species Distributions: The Effects of Niche Model Performance and Niche Properties. *PLOS ONE*, **10**, e0120056.

Additional Supporting Information and appendices will be available when this pre-print becomes published in a peer-reviewed journal.